\documentclass[letterpaper,usenatbib]{mn2e}
\usepackage{graphicx}
\usepackage{color}
\usepackage{lscape}
\usepackage{rotating}
\usepackage{longtable}
\usepackage{epstopdf}
\newsavebox{\tablebox}
\usepackage{hyperref}
\usepackage{times}
\usepackage{longtable,lscape}

\newcommand{\ergs}{\ifmmode {\rm erg\ s}^{-1} \else erg s$^{-1}$\ \fi}

\newcommand{\feii}{Fe {\sc ii}\ }

\newcommand{\civ}{C {\sc iv}\ }

\newcommand{\lb}{\ifmmode L_{\rm Bol} \else $L_{\rm Bol}$\ \fi}
\newcommand{\ledd}{\ifmmode L_{\rm Edd} \else $L_{\rm Edd}$\ \fi}
\newcommand{\leddR}{\ifmmode L_{\rm Bol}/L_{\rm Edd} \else $L_{\rm Bol}/L_{\rm Edd}$\ \fi}
\newcommand{\lx}{\ifmmode L_{\rm 2-10keV} \else  $L_{\rm 2-10keV}$\ \fi}
\newcommand{\hb}{\ifmmode H\beta \else H$\beta$\ \fi}
\newcommand{\ha}{\ifmmode H\alpha \else H$\alpha$\ \fi}
\newcommand{\hg}{\ifmmode H\alpha \else H$\gamma$\ \fi}
\newcommand{\oiii}{[O {\sc iii}]\ }
\newcommand{\oii}{[O {\sc ii}]\ }

\newcommand{\heii}{He {\sc ii}\ }
\newcommand{\mbh}{\ifmmode M_{\rm BH}  \else $M_{\rm BH}$\ \fi}
\newcommand{\lv}{\ifmmode \lambda L_{\lambda}(1350\AA) \else $\lambda L_{\lambda}(1350\AA)$\ \fi}
\newcommand{\lcon}{\ifmmode L_{1350} \else $L_{1350}$\ \fi}

\newcommand{\mdot}{\ifmmode \dot{m} \else \dot{m} \fi }
\newcommand{\llog}{\ifmmode {\rm log} \else {\rm log} \fi }
\newcommand{\kms}{\ifmmode {\rm km\ s}^{-1} \else km s$^{-1}$\ \fi}
\newcommand{\aox}{$\alpha_{\rm ox}$}

\newcommand{\nev}{[Ne {\sc v}]\ }
\newcommand{\neii}{[Ne {\sc ii}]\ }
\newcommand{\neiii}{[Ne {\sc iii}]\ }
\newcommand{\Siii}{[S {\sc iii}]\ }
\newcommand{\oiv}{[O {\sc iv}]\ }
\newcommand{\siv}{[S {\sc iv}]\ }

\begin{document}
\title[SPCA of mid-IR spectra of PG QSOs]{Spectral principal component analysis of mid-infrared spectra of a sample of PG QSOs}
\author[Bian et al.]{Wei-Hao Bian$^{1}$\thanks{E-mail: whbian@njnu.edu.cn}, Zhi-Cheng He$^{1}$\thanks{E-mail: zhichengho@126.com}, Richard Green$^{2}$, Yong Shi$^{3}$, Xue Ge$^{1}$ and Wen-Shuai Liu$^{1}$ \\
$^{1}$ Department of Physics and Institute of Theoretical Physics, Nanjing
Normal University, Nanjing 210023, China\\
$^{2}$ Steward Observatory, University of Arizona, Tucson, AZ 85721, USA\\
$^{3}$ School of Astronomy and Space Science, Nanjing University, Nanjing 210093, China
}
\maketitle
\begin{abstract}
A spectral principal component analysis (SPCA) of a sample of 87 PG QSOs at $z < 0.5$ is presented for their mid-infrared spectra from Spitzer Space Telescope. We have derived the first five eigenspectra, which account for 85.2\% of the mid-infrared spectral variation. It is found that the first eigenspectrum represents the mid-infrared slope, forbidden emission line strength and $9.7~\mu m$ silicate feature; the 3rd and 4th eigenspectra represent the silicate features at $18~ \mu m$ and $9.7~\mu m$, respectively. With the principal components (PC) from optical PCA, we find that there is a medium strong correlation between spectral SPC1 and PC2 (accretion rate). It suggests that more nuclear contribution to the near-IR spectrum leads to the change of mid-IR slope. We find mid-IR forbidden lines are suppressed with higher accretion rate. A medium strong correlation between SPC3 and PC1 (Eddington ratio) suggests a connection between the silicate feature at $18~\mu m$ and the Eddington ratio. For the ratio of the silicate strength at 9.7 $\mu m$ to that at 18 $\mu m$, we find a strong correlation with PC2 (accretion rate or QSO luminosity). We also find that there is a medium strong correlation between the star formation rate (SFR) and PC2. It implies a correlation between star formation rate and the central accretion rate in PG QSOs.
\end{abstract}


\section{INTRODUCTION}

For active galactic nuclei (AGN) and QSOs, the standard paradigm posits an accretion disk surrounding a central supermassive black hole (SMBH), along with other components, such as the broad-line region (BLR), narrow-line region (NLR), jet, and torus \citep[e.g.,][]{UP95}. This paradigm is supported by observations in multiple wavelengths. The radiation from different components of AGN/QSOs, as well as the orientation, leads to the diversity of AGN/QSO spectral energy distributions. Multivariate correlation analysis is a typical method to reveal physical connections among the emitting components of AGN/QSOs.

Principal component analysis (PCA) is used in multivariate correlation analysis \citep[e.g.,][]{BG92, Sulentic00}. It is a powerful mathematical tool used to reduce the dimensionality of a data set, describing the variance in the data set by the fewest number of variables called principal components (PCs). PCA uses a set of measured input variables, which can be subject to systematic errors of measurement, and is model-dependent if physical variables are derived \citep[e.g.,][]{BG92}. Spectral principal component analysis (SPCA) is an alternative approach, which uses the flux in each wavelength bin as input variables \citep[e.g.,][]{Mittaz90, Francis92, Shang03, Yip04, BL10, Wang11, Paris11, Hu12, Lee12}. SPCA doesn't depend on measurements that might have systematic errors.

For a complete optically selected sample of low-redshift Palomar-Green (PG) QSOs \citep[$z<0.5$,][]{SG83}, \cite{BG92} used PCA and found that most of the diversity/variance in the optical emission-line properties and continuum properties (radio through X-ray) was contained in two sets of correlations, eigenvectors of the correlation matrix. Principal component 1 (PC1) links the strength of optical \feii emission, \oiii emission, and \hb line asymmetry. Principal component 2 (PC2) involves optical luminosity and the strength of \heii 4686 and \aox. With the single-epoch SMBH mass derived from \hb \citep{Kaspi00}, it is found that PC1 is driven predominantly by the Eddington ratio, and PC2 is driven by the accretion rate \citep{Boroson02}.

Recent advances in the study of normal galaxies and AGN provide ample observational evidence for the existence of central SMBHs and the relationship between SMBHs and bulge properties of host galaxies \citep[e.g.,][] {FM00, Gebhardt00, Tremaine02, Shen08, Shen15}. This coevolution of the SMBH and the host galaxy implies some causal connection between the AGN and star formation properties \citep[e.g.,][]{KH13}. Probing the properties of the dusty torus/host surrounding SMBHs should offer important clues to the supply for the accretion, and the feedback to maintain this correlation.

The infrared (IR) spectra of QSOs are rich in features, such as hot/cold dust emission, silicate features at 9.7 and 18 $\mu m$ in emission or absorption, polycyclic aromatic hydrocarbon (PAH) features mainly at 6.2, 7.7, 8.6, 11.3, and 12.7 $\mu m$, atomic fine-structure and molecular hydrogen emission lines  (H$_2$ $9.67~\mu m$, \siv $10.5~\mu m$, \neii $12.8~ \mu m$, \neiii $15.6~ \mu m$, H$_2$ $17.03~ \mu m$, \Siii $18.7~\mu m$, \nev $24.3~\mu m$, \oiv $25.9 ~\mu m$) \citep[e.g.,][]{Hao05, Hao07, Cao08, Smith07, Shi14}. The IR spectra are powerful in probing the dusty torus properties. Dust in the torus is heated by AGN UV/optical radiation from the accretion disk to high temperature ($\sim$ 100-1000 K). The radiation by the dusty torus is dominated by a featureless continuum, with broad silicate features at around 9.7 $\mu m$ and 18 $\mu m$ in emission or absorption, \nev at 14.3 $\mu m$ and 24.3 $\mu m$ \citep[e.g.,][]{Hao07,Shi14}. The difference in silicate features for QSOs/AGN/ultraluminous infrared galaxies (ULIRGs) suggested the smooth/clumpy torus \citep{Hao07, Feltre12, Shi14}.

IR data are also powerful in probing the host galaxy properties, e.g., the star formation rate (SFR) \citep[e.g.,][]{Smith07, Shi14}. The host PAH features dominate at wavelengths shorter than 15 $\mu m$, while at longer wavelengths there is a strong continuum due to warm and cold ($<60 K$) dust heated by the host star formation \citep[e.g.,][]{Smith07, Shi14}. For AGN/QSO, the commonly used SFR tracers, such as UV radiation, \ha, and \oii are contaminated severely by the nuclear radiation; the mid-IR PAH features (e.g., 11.3 $\mu m$) and far-IR photometry are two relatively uncontaminated tracers of SFR in quasars \citep[e.g.,][]{Shi14, Netzer07, Shi14, Gurkan15}. \cite{Wang11} presented mid-IR SPCA applied to a sample of 119 Spitzer spectra of ultra-luminous infrared galaxies (ULIRGs) at $z < 0.35$. They found that the first SPC is characterized by dust properties such as temperature and the geometry. The second SPC is about the SFR. The third SPC represents a mid-IR contribution from the AGN.

In this paper, we have derived the first five SPCs of mid-IR Spitzer spectra for a complete sample of 87 PG QSOs to investigate the characteristics of mid-IR spectra, as well as the relation with the optical PCs from optical spectra.  \S 2 describes data and \S 3 is about the SPCA. \S 4 contains results and discussion. The Summary is given in the last section. Throughout this work, we use a cosmology with $H_0 = 70 \kms \rm Mpc^{-1}$, $\Omega_M = 0.3$, and $\Omega_{\Lambda} = 0.7$.

\section{Data}

The complete optically selected sample of 87 PG QSOs at $z < 0.5$ is used as our sample \citep{SG83, BG92}. These PG QSOs are defined by a limiting B-band magnitude of 16.16, blue $U-B$ color ($<-0.44$), and dominant starlike appearance. All these PG QSOs show broad emission lines, and thus are classified as type 1 QSOs. This sample is representative of bright optically selected QSOs \citep{Jester05}. It is the most thoroughly explored sample of AGN, with a lot of high quality data at most wave bands \citep[e.g.,][]{BG92, Brandt00, BL04, Shi14}. \cite{BG92} obtained optical spectra covering the region $\lambda \lambda 4300-5700$ for all 87 PG QSOs, and presented PCA results (i.e., PCs). With updated data, the optical PCs were calculated in \cite{Boroson02}. \cite{BL04} used the UV spectra to investigate the \civ line for 81 PG QSOs. For the SMBH masses for PG QSOs, we adopted those of \cite{VP06}, where they used a new empirical R-L relation \citep{Kaspi05, Bentz06}.  With a bolometric correction (BC) from the luminosity at 5100 \AA\ ($L_{\rm 5100}$) of $BC_{5100}=9.0$, we calculate the Eddington ratio \leddR.

We use Spitzer Space Telescope Infrared Spectrograph (IRS) data ($5-40~\mu m$ in the observed frame) and MIPS photometric data for all 87 PG QSOs, which have been carefully reduced by \cite{Shi14}. In their Table 2, \cite{Shi14} gave the factor used to multiply with the IRS short-low to long-low modules (in Column 7), and the factor to scale the IRS to the MIPS $24 ~\mu m$ photometry (in Column 8). They included the contributions from the torus and star forming region (multi-temperature black-body, silicate features, aromatic features, emission lines, and the star-forming template) to model the rest-frame IRS spectra and broad-band photometry. SFR were estimated for PG QSOs by combining three methods (i.e., $11.3 ~\mu m$
polycyclic aromatic hydrocarbon (PAH) feature, MIPS $70~\mu m$ photometry, and MIPS/Herschel PACS $160 ~\mu m$ photometry) through template fitting. The results for measured features and SFR were presented in Table 3 in \cite{Shi14}.

\section{Spectral PRINCIPAL COMPONENT ANALYSIS (SPCA)}

\subsection{A brief introduction to SPCA}
Instead of using measured variables, the flux in each wavelength bin is used in the SPCA as the input variable in searching for correlations. A primary benefit of SPCA arises from the quantification of the importance of each eigenvector for describing the variation in the spectral data set \citep[e.g.,][]{Mittaz90, Francis92, Wang11, Hu12}. For $M$ spectra of QSOs, we have a $N\times M$ matrix $G$, where N is the number of wavelength bins. The matrix $G$ has element $g_{ik}$ as the flux of the $k$th mean-subtracted spectrum over the standard deviation in the $i$th wavelength bin. We can then form a covariance matrix $C$ of all the spectra,
\begin{equation}
C_{ij}= \frac{1}{M}\sum_{k=1}^{k=M}{g_{ik}\times g_{jk}}, ~~~~~1\leq i,j\leq N,
\end{equation}
The goal of SPCA is to calculate the eigenvalues and eigenvectors of the covariance matrix $C$. The covariance matrix $C$ can be diagonalized,
\begin{equation}
C=EDE^{T},
\end{equation}
where D is a diagonal matrix and its diagonal terms are $N$ eigenvalues ($\lambda_i$) of the corresponding $N$ eigenvectors (or eigenspectra) $E_i=[e_{i1}, e_{i2},..., e_{iN}]$. The eigenspectra are called spectral principal components (SPCs), each of which is a linear combination of the original spectra. SPCs are ordered by their corresponding square eigenvalues. The proportion of the variation accounted for by the $k$th
eigenspectrum can be calculated as $\lambda_k^2/\sum_{j=1}^{j=N} \lambda_j^2$. In some cases, only a few SPCs are needed to describe the original spectral data set.

\subsection{Implementation of SPCA and the derived SPCs}

\begin{figure}
\begin{center}
\includegraphics[height=9.0cm,angle=-90]{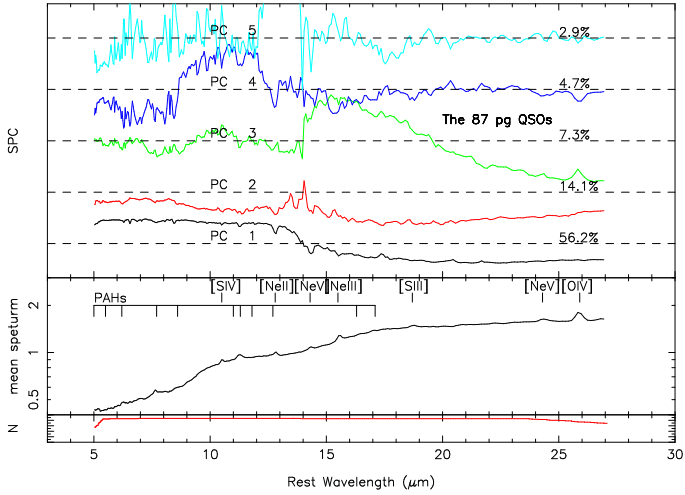}
\caption{The mean spectrum (bottom panel) and first five spectral principal components (top panel) of infrared spectra of 87 PG QSOs, as well as their contributions to the spectral variation. In the bottom panel, some emission/absorption features are labeled, as well as the number of spectra.}
\label{fig1}
\end{center}
\end{figure}

\begin{figure}
\begin{center}
\includegraphics[height=7.0cm,angle=-90]{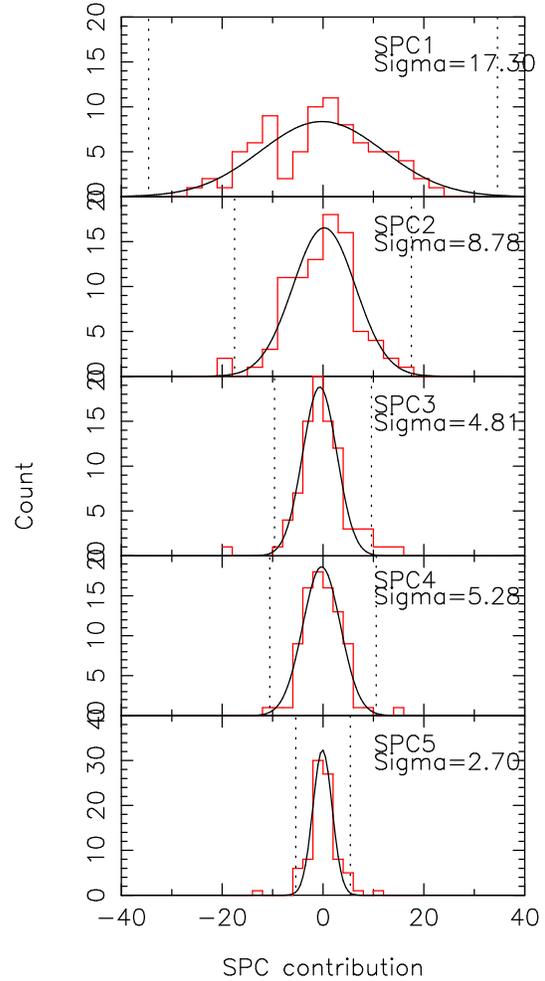}
\caption{Histogram of contributions to each of the 87 PG QSO spectra from each
of the first five SPCs. The black curves are Gaussian fits to the histograms. The standard
deviation of each Gaussian fit is shown in each panel. The dashed lines
mark the $2\sigma$ range. A greater width of the Gaussian distribution indicates
that more sources need a contribution from that particular SPC. }
\label{fig2}
\end{center}
\end{figure}

First of all, the observed spectrum is corrected to the rest frame by its redshift. We chose a wavelength range from 5.0 $\mu m$ to 27.0 $\mu m$ in the rest frame, which covers most of the spectra in our data set. We divide this range into 318 equally spaced bins in logarithmic wavelength space. Each spectrum is normalized at 14 $\mu m$, and then the mean and standard deviation are calculated at each wavelength bin. We form a covariance matrix of all mean-subtracted spectra and then derive the eigenvectors/SPCs.

The first five SPCs and the mean spectrum are shown in Fig. ~\ref{fig1}. The SPCs represent the difference between the observed spectra and the mean spectrum of the sample. In the mean spectrum, we can clearly see features such as broad PAH emissions, neon fine-structure lines, silicate features at around $18 ~\mu m$ and $9.7 ~\mu m$. From Fig. \ref{fig1}, it is found that SPC1 and SPC2 contain strong variance from the spectral slope, the forbidden emission lines and the 9.7 $\mu m$ silicate feature. SPC3 and SP4 reflect the variance particularly by the silicate features at $18 ~\mu m$ and $9.7 ~\mu m$, respectively.

The $i$th SPC value for the $j$th PG QSO is given by the linear combination of the normalized variables \citep[e.g.,][]{Francis92, Boroson02, Hu12}:
\begin{equation}
{\rm SPC}_{ij}=\sum_{k=1}^{k=N} e_{ik}\times \frac{f_{kj}-\mu_{k}}{\sigma_{k}}
\end{equation}
where $e_{ik}$ is the coefficient for the $i$th principal component and the $k$th spectral wavelength bin,
$f_{kj}$ is the spectral data for the $k$th spectral wavelength bin of the $j$th PG QSO,
and $\mu_{k}$ and $\sigma_{k}$ are the mean and standard deviation of the $k$th spectral
wavelength bin.

The histogram of $SPC_{ij}$ is plotted in Fig.~\ref{fig2}. In each panel, the distribution can be fitted by a Gaussian function, the width of which decreases from SPC1 to SPC5 as expected. The width of SPC5 is significantly narrower than the first four SPCs, meaning that the number of objects that need significant contributions from SPC5 is small  \citep[e.g.,][]{Wang11}.

\section{RESULTS AND DISCUSSION}

\subsection{The first 5 mid-IR eigenspectra}

\begin{figure*}
\begin{center}
\includegraphics[height=15.0cm,angle=-90]{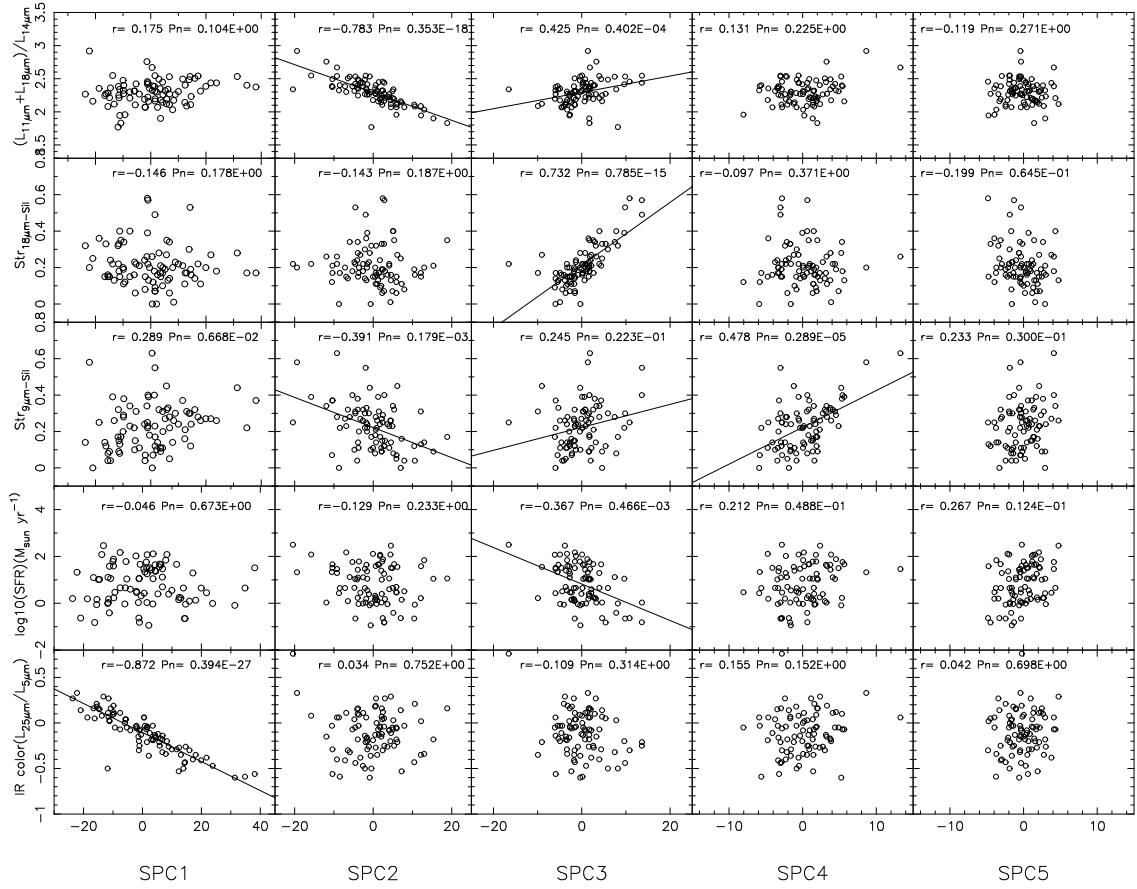}
\caption{SPC1-5 versus the IR color ($L_{25\mu m}/L_{5\mu m}$), SFR, the $9.7\mu m$ silicate feature, the $18\mu m$ silicate feature, $(L_{11\mu m}+L_{18\mu m})/L_{14\mu m}$ (from bottom to top). SPC1 has a strong correlation with mid-IR slope, $L_{25\mu m}/L_{5\mu m}$. SPC2 has a strong correlation with $(L_{11\mu m}+L_{18\mu m})/L_{14\mu m}$. SPC3 has a strong correlation with the silicate feature at $18 \mu m$ and a medium strong correlation with the silicate feature at $9.7 \mu m$ and SFR. SPC4 has a medium strong correlation with the silicate feature at $9.7\mu m$.}
\label{fig3}
\end{center}
\end{figure*}

\begin{table*}
\centering
\caption{Summary of the Spearman correlation coefficients between infrared SPCs, and IR parameters, such as IR color ($L_{25\mu m}/L_{5\mu m}$), SFR, $9.7\mu m$ silicate feature, $18\mu m$ silicate feature, $(L_{11\mu m}+L_{18\mu m})/L_{14\mu m}$. The correlations between these IR parameters and the accretion parameters (\mbh,\leddR,$L_{\rm 5100}$, PC1, PC2) are also shown. The value in brackets is the probability of the null hypothesis. The significant/medium strong correlations are shown in bold.}
\label{table1}
\begin{tabular}{lccccccccccccccccccc}
\hline
     & $L_{25\mu m}/L_{5\mu m}$  &  SFR  & $Str_{9.7\mu m}$    &$Str_{18\mu m}$    & $(L_{11\mu m}+L_{18\mu m})/L_{14\mu m}$  \\
\hline
SPC1 &\textbf{-0.87(3.9E-28)}    & -0.05(0.7)     &0.29(6.7E-03)     & -0.15(0.18)        & 0.18(0.1)      \\
SPC2 &0.03(0.75)      & 0.13(0.23)        &  \textbf{-0.39(1.8E-04)} &  -0.14(0.19)     & \textbf{-0.78(3.5E-19) }      \\
SPC3 & -0.1(0.31)     & \textbf{-0.37(4.7E-04) }  & 0.25(0.02)      &  \textbf{0.73(7.9E-16)}     & \textbf{0.43(4.0E-05) }    \\
SPC4 & 0.16(0.15)    & 0.2(0.05)        &  \textbf{0.48(2.9E-06)}   &  0.1(0.4)        & 0.13(0.23)        \\
SPC5 &0.04(0.7)      &   0.27(0.01)     & 0.23(0.03)     & -0.2(0.06)        & -0.12(0.27)    \\
\hline
\mbh & -0.20(6.3E-2)  & \textbf{0.50(8.8E-7) }    & \textbf{0.46(8.5E-6)}   &0.05(0.67)         & 0.14(0.18)
\\
\leddR & 0.05(0.62)  & 0.28(8.4E-3)     &-0.10(0.35)     &-0.32(2.5E-3)      & -0.13(0.23)
\\
$L_{\rm 5100}$ & -0.19(7.8E-2) & \textbf{0.77(2.6E-18)} & \textbf{0.46(9.1E-06)} & -0.16(0.13) & 0.10(0.33)\\
PC1 &  -0.00 (-0.98) & 0.29(6.6E-3) & 0.32(2.3E-3) & \textbf{0.41(7.2E-05)} & 0.25(0.02) \\
PC2 &  0.28(8.38E-3) & \textbf{-0.61(2.7E-10)} & \textbf{-0.40(1.0E-04)} & 0.27(0.01) & -0.03(0.79)\\
\hline
\label{t1}
\end{tabular}
\end{table*}

\begin{figure}
\begin{center}
\includegraphics[height=9.0cm,angle=-90]{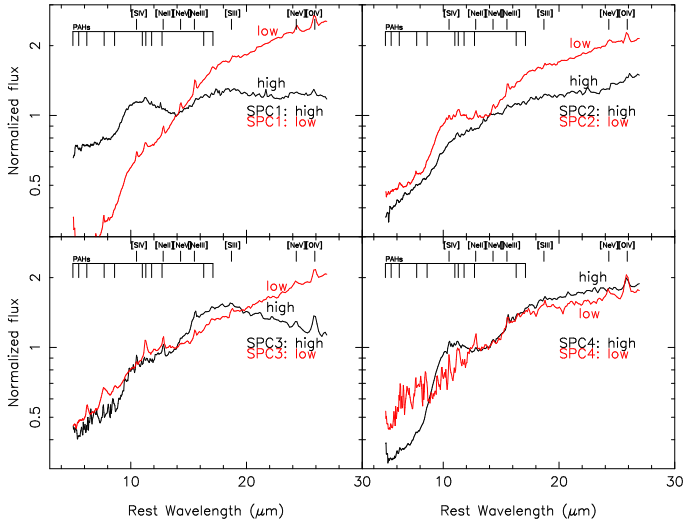}
\caption{Ilustration of the range of variation contained in the spectral eigenvectors.  The two traces in each figure show the mean spectrum divided by an SPC (one is $\rm SPC > 1\sigma$ in the SPC distribution in Fig. \ref{fig2}, another is $\rm SPC < -1\sigma .$: upper left is for SPC1; upper right is for SPC2; bottom left is for SPC3; bottom right is for SPC4.) Some emisison/absorption features are labeled.}
\label{fig4}
\end{center}
\end{figure}

In Fig.~\ref{fig1}, the weight decreases from SPC1 to SPC5 as expected, and the sum of the weights of the first five SPCs is 85.2\%. In Table ~\ref{t1}, we give the Spearman correlation coefficient $R$ and the probability of the null hypothesis $P_{\rm null}$ for the correlations between SPCs 1-5 and five parameters, i.e., the IR color ($L_{25\mu m}/L_{5\mu m}$), SFR, the $9.7\mu m$ silicate feature, the $18\mu m$ silicate feature, and $(L_{11\mu m}+L_{18\mu m})/L_{14\mu m}$.

In Fig.~\ref{fig3}, we also show the best linear fits for the strong and medium strong relations shown in Table ~\ref{t1}. For SPC1, there is a strong correlation with the mid-IR slope $L_{25\mu m}/L_{5\mu m}$ with $R=-0.87, P_{\rm null} = 3.9E-28$. \cite{Cao08} found that colour index $\alpha(30,15)$ is a good indicator of the relative contribution of starbursts to AGN mid-IR spectra. It implied the connection between IR slope and AGN accretion rate, although the correlation between them is weak (Table ~\ref{t1}).

For SPC2, there is a strong correlation with a measure of the silicate feature and PAH strength $(L_{11\mu m}+L_{18\mu m})/L_{14\mu m}$ with $R=-0.78, P_{\rm null} = 3.5E-19$. SPC2 has a medium strong correlation with the silicate feature at 9.7 $\mu m$, $Str_{9.7\mu m}$,  ($R=-0.39, P_{\rm null} = 1.8E-4$) and PAH EW at 11.3 $\mu m$, $EW_{11.3\mu m}$, ($R=-0.24, P_{\rm null} = 2.5E-2$). SPC3 has a strong correlation with the silicate emission at $18~\mu m$ ($R=0.73, P_{\rm null} = 7.9E-16$). SPC3 also has a medium strong correlation with the silicate/PAH measure $(L_{11\mu m}+L_{18\mu m})/L_{14\mu m}$  ($R=0.43, P_{\rm null} = 4.0E-5$), as well as with the SFR ($R=0.39, P_{\rm null} = 1.8E-4$) (Fig.~\ref{fig3}). SPC4 has a medium strong correlation with the silicate emission at $9.7~\mu m$ ($R=-0.48, P_{\rm null} = 2.9E-6$) (Fig. ~\ref{fig3}). We can't find any significant correlation for SPC5, which is consistent with the narrow distribution in Fig. \ref{fig2}.

For PG QSOs with extreme SPC1 values in the SPC1 distribution ($\rm |SPC1|>1~\sigma$) in Fig.~\ref{fig2}, we display their mean spectrum, which is given in the top left panel in Fig.~\ref{fig4}. It shows an obvious difference from the opposite extreme of the distribution for the mid-IR slope, with larger slope for smaller values of SPC1. That difference is consistent with the above interpretation of SPC1. SPC1 seems to contain the strength of the forbidden lines, such as \siv $10.5~\mu m$, \neii $12.8~ \mu m$, \neiii $15.6~ \mu m$, \nev $24.3~\mu m$, \oiv $25.9 ~\mu m$. The steeper the mid-IR slope, the stronger the lines. That implies a correlation between SPC1 and the strength of these mid-IR forbidden lines. For PG QSOs with higher values of SPC1, we also find that there appears to be a strong broad feature at about 10 $\mu m$ (the left top panel in Fig.~\ref{fig4}). It can be interpreted as a major, single-temperature dust component near the sublimation temperature of the smallest grains. We find a weak correlation between SPC1 and the silicate feature at 9.7 $\mu m$ (Table ~\ref{t1}), suggesting that broad feature at about 10 $\mu m$ also has a contribution from the silicate feature at 9.7 $\mu m$.

Fig. ~\ref{fig4} also shows other cases for SPC2, SPC3, SPC4. There exists an obvious silicate difference at $18~\mu m$, $9.7~\mu m$, respectively for SPC3, SPC4. We also find an obvious PAH emission at 11.3 $\mu m$ with lower values of SPC3 in the left bottom panel in Fig. ~\ref{fig4}. It is consistent with the medium strong correlation between SPC3 and the SFR (Table ~\ref{t1}).

\subsection{The relation with PCs by Boroson \& Green (1992)}

\begin{table*}
\centering
\caption{Summary of the Spearman correlation coefficients between infrared SPCs and optical PCs, as well as \mbh, \leddR. The value in brackets is the probability of the null hypothesis. The significant/medium strong correlations are shown in bold. }
\label{table1}
\begin{tabular}{lccccccccccccccccccc}
\hline
     &  PC1         &  PC2           & PC3            & PC4             & \mbh         & \leddR \\
\hline
SPC1 &-0.11(0.3)    & \textbf{-0.44(1.7E-05)} &\textbf{-0.38(2.6E-04)} & 0.14(0.2)        & 0.21(0.05)  & 0.06(0.57)      \\
SPC2 &-0.2(0.06)    & 0.09(0.4)      &  -0.09(0.4)   &  -0.1(0.34)      & -0.15(0.15)  & 0.004(0.97)      \\
SPC3 & \textbf{0.42(5.5E-05)}& 0.25(0.02)     & 0.07(0.53)    &  0.006(0.95)     & 0.1(0.35)     & \textbf{-0.42(6.4E-05)}\\
SPC4 & 0.02(0.89)   & -0.33(2.0E-03) &  0.12(0.26)   &  0.001(0.99)     & 0.26(0.01) & -0.01(0.92)        \\
SPC1+SPC4 &-0.08(0.44)&\textbf{-0.50(7.7E-07)} & -0.33(1.6E-03)& 0.14(0.19)      & 0.28(9.6E-03)& 0.04(0.71)       \\
\hline
\label{t2}
\end{tabular}
\end{table*}

With 13 properties for PG QSOs, \cite{BG92} presented their PCA and gave optical PC1 and PC2. Considering the central SMBH accretion process, \cite{Boroson02} presented a PCA by using updated data and found that PC1 is driven predominantly by the Eddington ratio \leddR and PC2 is driven by the accretion rate (proportional to \lb). Using Spearman correlation between our infrared spectral SPC values and optical PC values from \cite{BG92} (Table ~\ref{t2}), we want to discover any relation of the mid-IR SPCs with the central accretion process. In Table ~\ref{t1}, we also shown the relations between five IR parameters and the accretion parameters (\mbh,\leddR,$L_{\rm 5100}$, PC1, PC2).

For SPC1, which is characterised by the mid-IR spectral slope, we find that SPC1 has a medium strong correlation with optical PC2 from \cite{BG92} ($R=-0.44, P_{null} = 2.0E-05$, Table ~\ref{t2}). Because the optical PC2 is driven by the accretion rate \cite{Boroson02}, the above relation suggests that SPC1 (mid-IR spectral slope + emission features) is related to the central accretion process.

 There is no significant correlation between SPC1 and PC1 ($P_{null}=0.3$). We find that there is a weak correlation between SPC1 and \mbh ($R = 0.21$, $P_{null}$ = 0.05, Table ~\ref{t2}). It is also found that there is a weak correlation between SPC1 and Peak 5007 ($R=-0.37, P_{null} = 4.0E-04$). In the top left panel in Fig. \ref{fig4}, the forbidden lines are stronger for PG QSOs with steeper mid-IR slopes, implying a correlation between SPC1 and the strength of these mid-IR forbidden lines. Therefore, it is not surprising that SPC1 correlates with the optical forbidden line \oiii $\lambda$ 5007. The mid-IR/optical forbidden line strength seems to be in the same SPC1 as that major slope difference (left top panel in Fig. \ref{fig4}). Considering the medium strong correlation between SPC1 and optical PC2 (i.e., the accretion rate), it suggests that EWs for mid-IR/optical forbidden lines are suppressed because of a higher accretion rate (high short wavelength continuum, a flatter slope) \citep{Dietrich02}.

\begin{figure}
\begin{center}
\includegraphics[height=8cm,angle=-90]{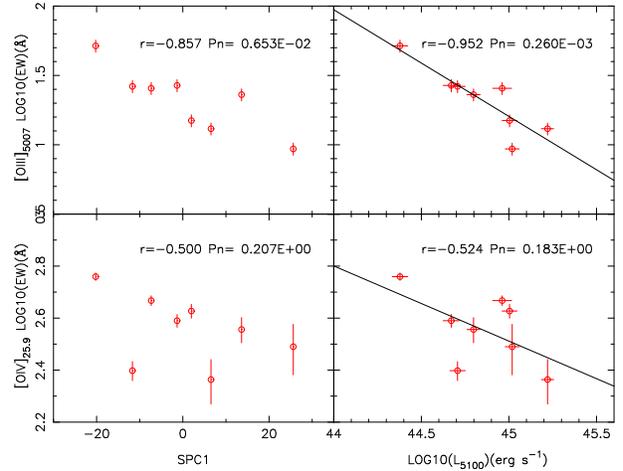}
\caption{Left: EW of optical \oiii (top) and \oiv (bottom) versus SPC1. Right: EW of \oiii (top) and \oiv (bottom) versus $L_{\rm 5100}$. In the right two panels, we also show the best linear fits as solid lines. The correlation coefficients and $P_{\rm null}$ are shown in the panels. }
\label{fig5}
\end{center}
\end{figure}

Because of the weakness of IR forbidden lines in IR spectra, we produced composite spectra in 8 bins based on SPC1 (about 10 QSOs per bin). We measure EWs for forbidden lines of \siv $10.5~\mu m$, \neii $12.8~ \mu m$, \nev $14.3~\mu m$, \neiii $15.6~ \mu m$, \nev $24.3~\mu m$, and \oiv $25.9 ~\mu m$. We find that there are anti-correlations between the EWs of the IR forbidden lines (\oiv $25.9 ~\mu m$, \nev $14.3~\mu m$) and SPC1 or $L_{\rm 5100}$. Considering the largest EW of \oiv $25.9 ~\mu m$, we take it as an example shown in Fig.~\ref{fig5}. The anti-correlation between EW of \oiv $25.9 ~\mu m$ and SPC1 is consistent with Fig.~\ref{fig4}. The anti-correlation between EW of IR \oiv $25.9 ~\mu m$  and $L_{\rm 5100}$ is similar to that between EW of optical \oiii 5007  and $L_{\rm 5100}$ \citep[e.g.,][]{Shen15b}. Increasing an order of magnitude in $L_{\rm 5100}$, the EW is decreased by about 0.77 dex, 0.29 dex for optical \oiii and IR \oiv $25.9 ~\mu m$, respectively. There is a relation between EW of optical \oiii and EW of IR \oiv $25.9 ~\mu m$ ($R=0.55, P_{null} = 0.16$), $EW(\rm [O ~IV])=(225.7\pm 66.4) +(6.28\pm 2.48) EW(\rm [O~III])$. The EW of IR \oiv is suppressed with higher accretion rate ($L_{\rm 5100}$), i.e., Baldwin effect. We find that there is a medium correlation between the silicate feature at $9.7\mu m$ and \mbh or $L_{\rm 5100}$ (Table ~\ref{t1}). It suggests more dust for QSOs with larger accretion rates. More dust could be related to a smaller opening angle for the exciting Optical/UV radiation, thus making a smaller NLR with lower forbidden line emission.

For SPC2, we can't find a significant correlation with the optical PCs (Table ~\ref{t2}). For SPC3, we find that there is a medium strong correlation with optical PC1 from \cite{BG92} ($R=-0.42, P_{null} = 5.5E-05$, Table ~\ref{t2}). Because the optical PC1 is driven by \leddR \citep{Boroson02}, this medium correlation between SPC3 and PC1 also shows the connection between SPC3 and the central accretion process. It is confirmed by a medium strong correlation between SPC3 and \leddR ($R=-0.42, P_{null} = 6.4E-05$). It is also found that there is a medium strong correlation between SPC3 and Peak 5007 ($R=0.52, P_{null} = 3.3E-07$). Considering the correlation between SPC3 and PC1/SFR, SPC3 is an important eigenspectrum for the connection with the central SMBH accretion and the star formation process.

It is found that there is a weak correlation between SPC4 and optical PC2 ($R=-0.33, P_{null} = 2.0E-03$, Table ~\ref{t2}). We also find that there is a weak correlation between SPC4 and \mbh ($R = 0.26, P_{null} = 0.01$, Table ~\ref{t2}). Considering both SPC1 and SPC4 have correlations with the optical PC2/\mbh, we add SPC4 to SPC1 as a new eigenspectrum. It is found the above two correlations are improved, with $R=-0.5, -0.28$ (Table ~\ref{t2}), respectively. We can not find SPC5 to have a significant correlation with any of the optical or IR parameters.

\begin{figure}
\begin{center}
\includegraphics[height=8cm,angle=-90]{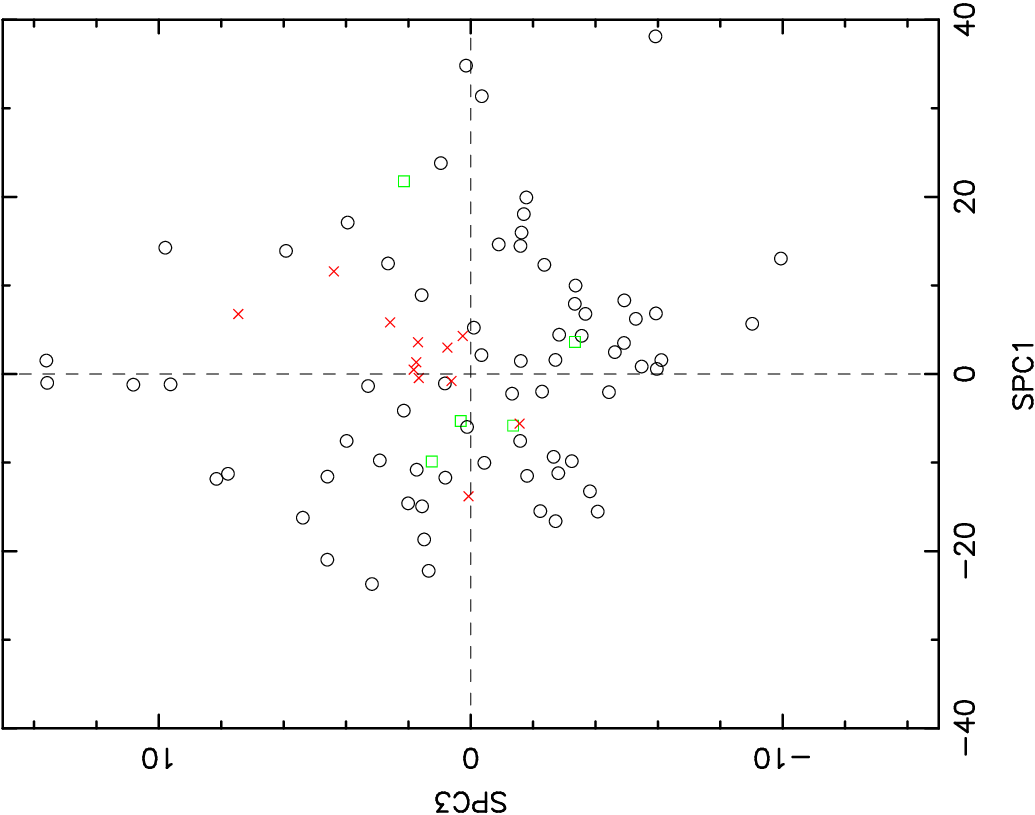}
\includegraphics[height=8cm,angle=-90]{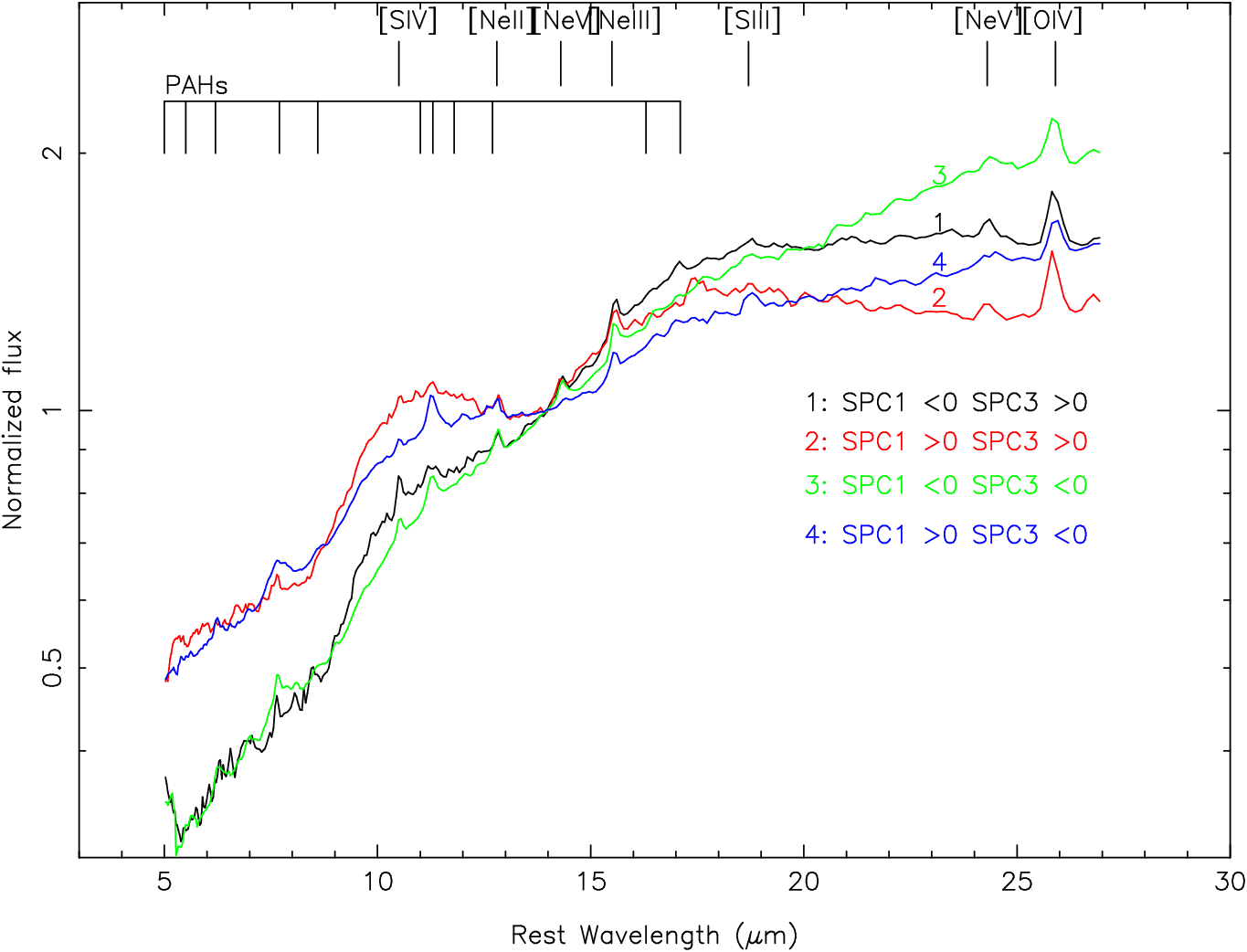}
\caption{Top: Distribution of the 87 low-redshift PG QSOs with respect to SPC1 versus SPC3. The circles denote radio-quiet QSOs. Crosses denote the radio-loud steep-spectrum QSOs, and squares are radio-loud flat-spectrum QSOs.  Bottom: Four mean spectra of the PG QSOs in the four quadrants, respectively.}
\label{fig6}
\end{center}
\end{figure}

Considering the medium strong correlation between SPC1 and PC2, as well as the correlation between SPC3 and PC1, we show the AGN contribution in the mid-IR spectra for PG QSOs through SPCA. The top panel in Fig.~\ref{fig6} shows the diagram of SPC1 versus SPC3. The bottom panel in Fig.~\ref{fig6} shows the mean spectra in the four quadrants, respectively. The mid-IR slope is different in the mean spectrum for PG QSOs with $\rm SPC1 > 0$ from that for PG QSOs with $\rm SPC1 < 0 $. The silicate difference at 18 $\mu m$ is also obvious for PG QSOs with $\rm SPC3 > 0 $. These results are consistent with Fig. ~\ref{fig3}. Considering the medium strong correlation between SPC1 and PC2 (accretion rate), it suggests a relation between the near-IR slope with the accretion rate. We find $R=-0.2, P_{\rm null}=0.06$ between the near-IR slope and SMBH mass.
For PG QSOs with larger SMBH mass (luminosity), more UV/optical photons irradiate the NLR along with less very hot dust. It would lead to a flat mid-IR spectrum.

The medium strong correlation between SPC3 and PC1 (the Eddington ratio) implies a connection between the silicate feature at 18 $\mu m$ and the Eddington ratio. We find that $R=-0.32, P_{\rm null}=2.5E-03$ between them (Table ~\ref{t1}). In the diagram of SPC1 versus SPC3, we note that most of the radio-loud steep-spectrum QSOs are located in the first quadrant, like the case that most of the radio-loud steep-spectrum QSOs are in the fourth quadrant in the diagram of optical PC1 versus PC2 in Fig. 1 in \cite{BG92}.

\begin{figure}
\begin{center}
\includegraphics[height=8cm,angle=-90]{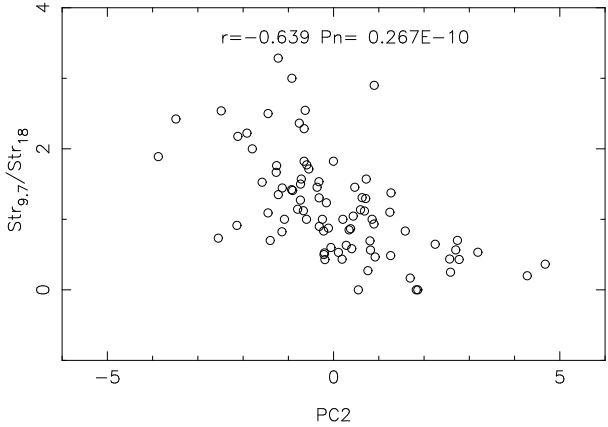}
\caption{$\rm Str_{9.7 \mu m}/Str_{18 \mu m}$ versus PC2. There is a strong correlation between this ratio and PC2 (accretion rate or QSO luminosity). The Spearman correlation coefficient R is 0.64 with $P_{\rm null} = 2.7\times 10^{-11}$.}
\label{fig7}
\end{center}
\end{figure}

Considering the ratio of the silicate strength at 9.7 $\mu m$ to that at 18 $\mu m$, we find a strong correlation between this ratio and PC2 with $R=-0.64, P_{\rm null}=2.7E-11$ (Fig. ~\ref{fig7}). For the relation between this ratio and \lb, $R=0.64, P_{\rm null}=3.7E-11$. For the relation between this ratio and the SMBH mass, $R=0.44, P_{\rm null}=2.4E-5$. The smaller R for the relation with the SMBH mass is possibly due to the uncertainties in SMBH masses (e.g., radio-loud QSOs, starlight contribution.).

\subsection{The connection between SFR and PCs}
\begin{figure}
\begin{center}
\includegraphics[height=8cm,angle=-90]{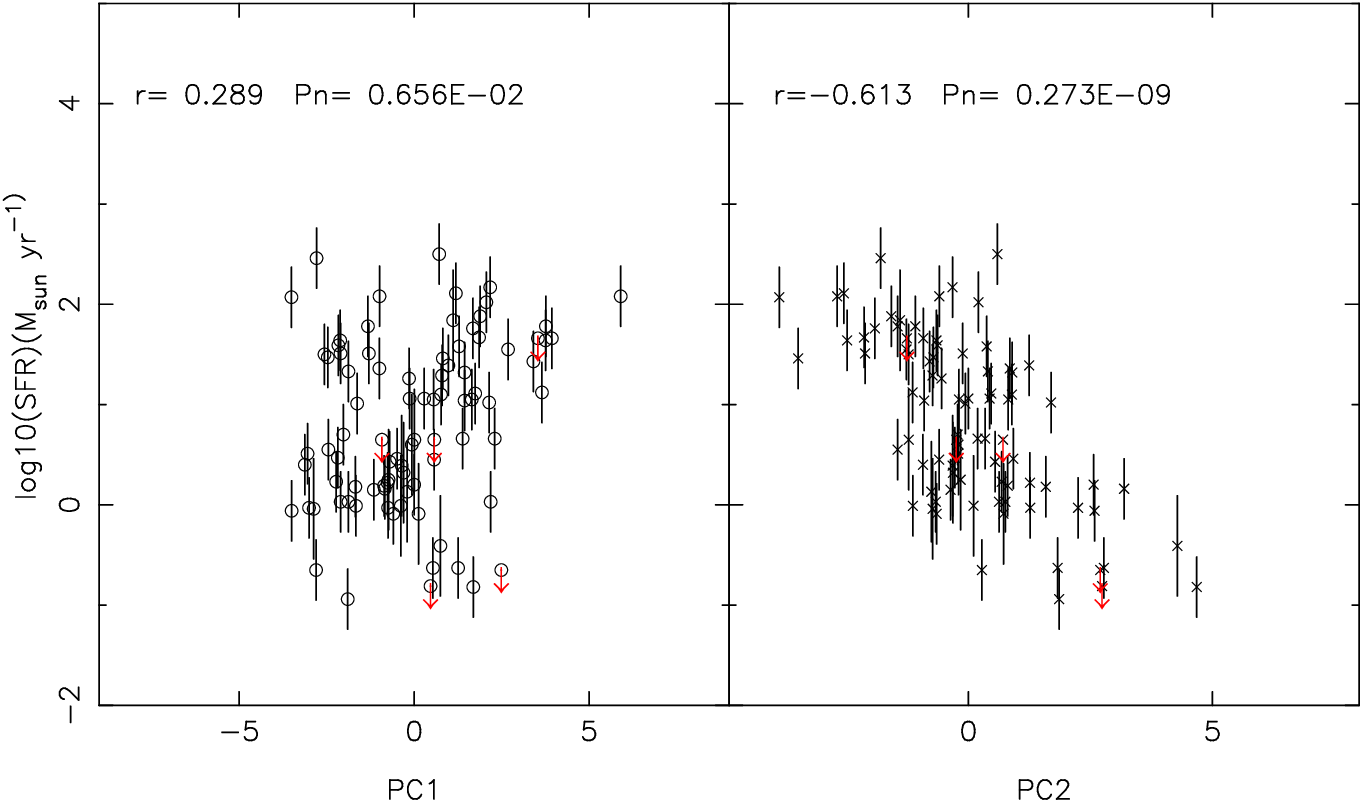}
\includegraphics[height=8cm,angle=-90]{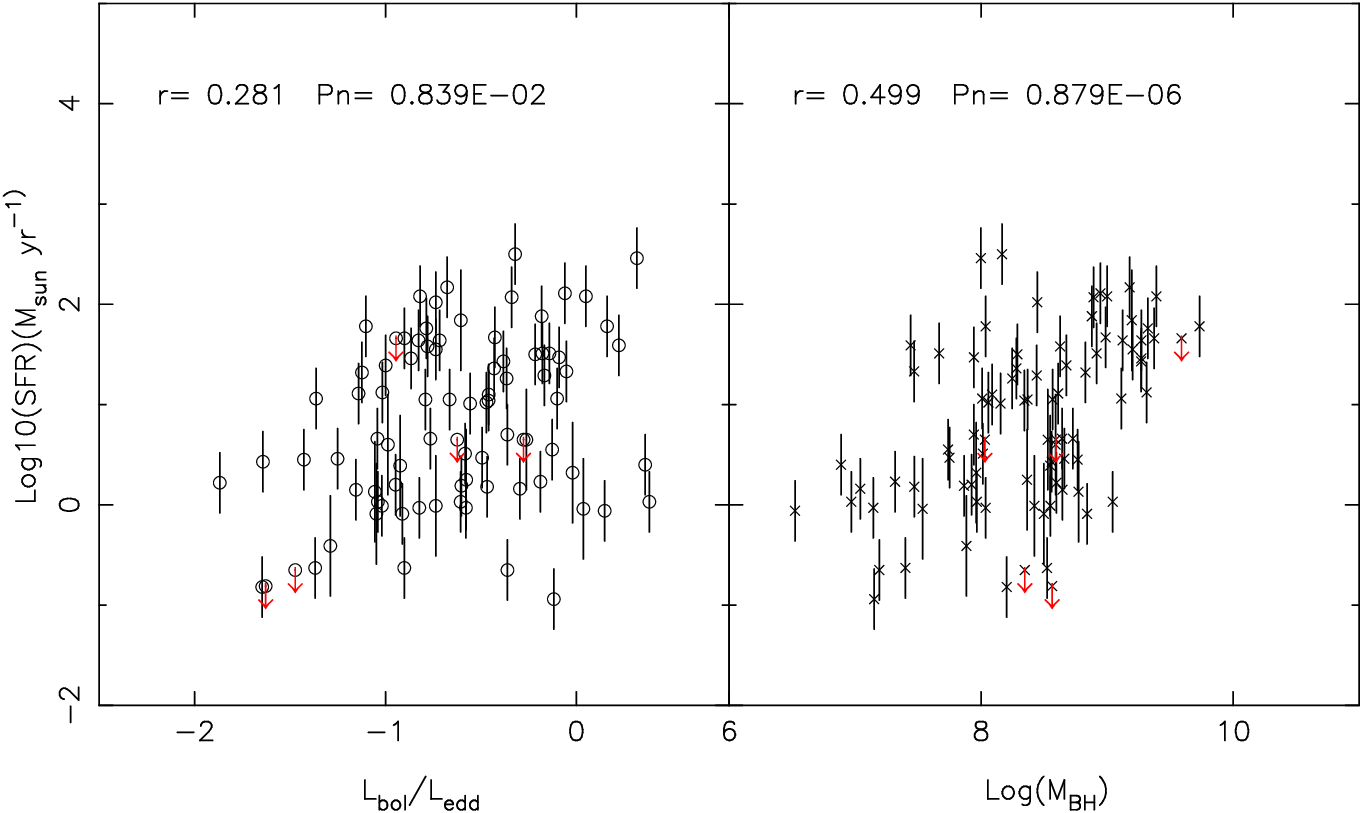}
\caption{The SFR of the host galaxies versus optical PC1, PC2, \leddR and \mbh. The SFR has a medium strong correlation with PC2 and \mbh (right panels). The SFR has a weak correlation with PC1 and \leddR (left panels).  The red arrows denote upper limits for the SFR. The correlation coefficients and $P_{\rm null}$ are shown in the panels.}
\label{fig8}
\end{center}
\end{figure}

With mid-IR spectra and/or far-IR photometry, \cite{Shi14} carefully derived the SFR for these PG QSOs through template fitting. The left panels in Fig. ~\ref{fig8} show the SFR \citep{Shi14} versus PC1  (Top) and \leddR (Bottom). The SFR has a weak correlation with both PC1 and \leddR , with $R=0.29$ ($P_{null}$ = 6.6E-03), and $R=0.28$ ($P_{null}$ = 8.4E-03), respectively. The right panels of Fig. ~\ref{fig8} show the SFR versus PC2 (Top) and \mbh (Bottom). The SFR has a medium strong correlation with optical PC2 and \mbh, with $R=-0.61$ ($P_{null}$ = 2.7E-10) and $R=0.50$ ($P_{null}$ = 8.8E-07), respectively. Excluding the redshift effect on PC2 and \mbh, we find that SFR still has a weak correlation with PC2 ($R = -0.176$ at $2.75 \sigma$ level), and a weak correlation with \mbh ($R=0.12$ at $2.0 \sigma$ level). It means that there is a certain connection between the SFR of the host galaxy and the central SMBH accretion process. The SFR has a strong correlation with $L_{\rm 5100}$ ($R=0.77, P_{\rm null} = 2.6E-18$, Table ~\ref{t1}). It is consistent with the previous work \citep[e.g.,][]{Hao05, Gurkan15}. It implies possible feedback, leading star formation in the host galaxy, by the central SMBH accretion process.

\section{SUMMARY}
A SPCA of mid-IR spectra from Spitzer Space Telescope ($5-40~\mu m$) is presented for a complete sample of 87 PG QSOs at $z < 0.5$. We have derived the first five eigenspectra, which account for 85.2\% of the total variation in the mid-IR spectra for PG QSOs. Their interpretation is explored through the SPC values in different wavelength bins, as well as the correlations with five measurement parameters. We also use the optical PCs from \cite{Boroson02} to investigate the connection of mid-IR spectra with the central SMBH accretion process. The main conclusions can be summarized as follows.

(1) SPC1 accounts for 56.2\% of the mid-IR spectral variation. It is found that SPC1 has a strong correlation with the mid-IR slope $L_{25\mu m}/L_{5\mu m}$. SPC1 seems to mark the strength of the mid-IR forbidden lines. The steeper the mid-IR slope, the stronger the forbidden lines. With the composite spectra in 8 bins of SPC1, we find the EW of \oiv $25.9\mu m$ or \nev $14.3\mu m$ is suppressed with higher accretion rate ($L_{\rm 5100}$), i.e., Baldwin effect. We also find that there is a medium correlation between the silicate feature at 9.7 $\mu m$ and \mbh or $L_{\rm 5100}$ (Table ~\ref{t1}). It suggests more dust for QSOs with larger accretion rates. More dust could be related to a smaller opening angle for the exciting Optical/UV radiation, thus making a smaller NLR with lower forbidden line emission. SPC3 represents the $18~\mu m$ silicate feature strength and the SFR. SPC4 has a medium strong correlation with the $9.7~\mu m$ silicate feature strength. For the diagram of SPC1 versus SPC3, the most obvious difference of four mean spectra is their mid-IR slope, as well as the silicate features. We note that most of the radio-loud steep-spectrum QSOs are located in the first quadrant in the diagram of SPC1 versus SPC3.

(2) Compared with the optical PCs by \cite{Boroson02}, we find that SPC1 has a medium strong correlation with optical PC2. SPC3 has a medium strong correlation with optical PC1. SPC4 has a weak correlation with optical PC2. The sum of SPC1 and SPC4 (SPC1+SPC4) has a stronger correlation with optical PC2. Considering the medium strong correlation between SPC1 and PC2, as well as the correlation between SPC3 and PC1, we show the AGN contribution in the mid-IR spectra through SPCA. A medium strong correlation between SPC3 and the PC1/Eddington ratio suggests a connection between the silicate feature at $18~\mu m$ and the Eddington ratio. For the ratio of the silicate strength at 9.7 $\mu m$ to that at 18 $\mu m$ ratio, we find a strong correlation with PC2 (accretion rate or QSO luminosity).

(3) Considering the SFR derived from the mid-IR by \cite{Shi14}, we find that there is a medium strong correlation between the SFR and optical PC2/\mbh, a weak correlation between the SFR and PC1 and \leddR. It is possible that the central accretion has a positive effect on the SFR of host galaxy. It means that there is a certain connection between the SFR of the host galaxy and the central SMBH accretion process. It implies possible feedback, leading star formation in the host galaxy, by the central SMBH accretion process.

\section{ACKNOWLEDGMENTS}
We are grateful to Francis P. J. for instructive comments. This work has been supported by the National Science Foundation of China (nos. 11373024, 11173016, and 11233003).

\end{document}